\documentclass[prl,aps,twocolumn,superscriptaddress]{revtex4}
\usepackage{amssymb}
\usepackage{amsmath}
\usepackage{float}
\usepackage{bbm}
\usepackage{graphicx}
\usepackage[naturalnames]{hyperref}

\begin{document}
\title{Noise-resistant control for a spin qubit array}
\author{J.~P.~Kestner}
\affiliation{Department of Physics, University of Maryland Baltimore County, Baltimore, MD 21250, USA}
\affiliation{Condensed Matter Theory Center, Department of Physics, University of Maryland, College Park, MD 20742, USA}
\author{Xin Wang}
\affiliation{Condensed Matter Theory Center, Department of Physics, University of Maryland, College Park, MD 20742, USA}
\author{Lev S. Bishop}
\affiliation{Condensed Matter Theory Center, Department of Physics, University of Maryland, College Park, MD 20742, USA}
\affiliation{Joint Quantum Institute, University of Maryland, College Park, MD 20742, USA}
\author{Edwin Barnes}
\affiliation{Condensed Matter Theory Center, Department of Physics, University of Maryland, College Park, MD 20742, USA}
\author{S.~Das Sarma}
\affiliation{Condensed Matter Theory Center, Department of Physics, University of Maryland, College Park, MD 20742, USA}
\affiliation{Joint Quantum Institute, University of Maryland, College Park, MD 20742, USA}

\begin{abstract}
We develop a systematic method of performing corrected gate operations on an array of exchange-coupled singlet-triplet qubits in the presence of both fluctuating nuclear Overhauser field gradients and charge noise.  The single-qubit control sequences we present have a simple form, are relatively short, and form the building blocks of a corrected {\sc cnot} gate when also implemented on the inter-qubit exchange link.  This is a key step towards enabling large-scale quantum computation in a semiconductor-based architecture by facilitating error reduction below the quantum error correction threshold for both single-qubit and multi-qubit gate operations.
\end{abstract}

\maketitle

The prospective scalability of spin qubits in semiconductor quantum dots, along with their demonstrated long coherence times and rapid gate operations, make them a leading candidate for quantum computing architectures. Singlet-triplet qubits \cite{Petta05,Maune12}, where the quantum information is encoded in the zero-projection spin states of two electrons in a double quantum dot, are particularly promising due to their insensitivity to stray magnetic fields and their purely electrical controllability. However, precise experimental manipulation of singlet-triplet qubits is hindered by two sources of noise invariably present in all laboratory systems: fluctuations in the background nuclear spin bath due to long-range hyperfine-mediated flip-flop processes (Overhauser noise) \cite{Reilly10,Cywinski09,Barnes12}, and fluctuations in the electrostatic quantum dot confinement potential due to background electrons hopping on and off nearby impurity sites (charge noise) \cite{Hu06,Culcer09}. These fluctuations are slow ($\sim 100 \mu$s) compared to typical qubit rotation times ($\sim 0.1$ ns), and this highly non-Markovian characteristic can be exploited to suppress their effects by spin echo or related dynamical decoupling protocols when one does not aim to rotate but instead preserve the qubit state, i.e. quantum memory \cite{Bluhm11,Barthel10,Medford12}.  The ability to achieve very long quantum memory times in this way is one of the great advantages of semiconductor spin qubit architectures.  The problem is that the dynamical decoupling scheme to preserve coherence does not work during the qubit gate operations.

Similar protocols for robust qubit rotations are, therefore, highly desirable so that quantum coherence is preserved during the gate operations.  However, singlet-triplet qubit control is subject to severe physical constraints that make this task quite daunting.  A large inter-dot Overhauser field gradient introduced by nuclear spin pumping \cite{Foletti09,Bluhm10,Ludwig12} (or an actual magnetic field gradient due to a proximal micromagnet \cite{Brunner11}) performs rotations about the $x$ axis of the Bloch sphere, but this gradient is not tunable on the time scale of an operation.  Meanwhile, electrical control of the inter-dot exchange coupling by tilting the double-well potential \cite{Petta05} leads to rotation about the $z$ axis of the Bloch sphere, but the sign of the coupling is fixed.  The only rapidly tunable element is the magnitude of the exchange coupling, and its range is limited by either the singlet-triplet splitting of two electrons on a single dot, or, more restrictively, by keeping the double-well near a symmetric configuration to avoid converting the spin qubit into a fragile charge qubit. Thus, one is restricted to positive rotations about a limited range of axes somewhere between $+\hat{x}$ and $+\hat{z}$, and pulses approximating a delta function are not available.  This unique set of constraints prohibits straightforward application of control techniques from the NMR literature (e.g., \cite{Wimperis94,Khodjasteh09}).  The high-fidelity gate operations crucial to scalable quantum computation requires a totally new approach to the quantum control problem.

Previously, we have shown that there exists a new form of control sequence that respects these constraints and eliminates the leading-order single-qubit error due to Overhauser noise \cite{Wang12}.  Recent numerical work has shown that both relevant types of error can be simultaneously addressed \cite{Khodjasteh12}. Despite this progress, there remains a need for a protocol that corrects both Overhauser and charge noise errors while being sufficiently simple and flexible for incorporation into multi-qubit operations.

In this work, we present a method of pulse design that achieves this goal, systematically eliminating both errors to leading order for any quantum circuit.  In the language of the NMR community, the task is similar to correcting arbitrary quantum gate operations for both amplitude and detuning errors simultaneously.  This task has only recently begun to be addressed in NMR \cite{Bando12,Merrill12}, and it is remarkable that the far more restricted case of singlet-triplet qubits permits a simple solution, as revealed by our method.  Furthermore, our new pulse sequences are an order of magnitude faster than earlier, less-powerful sequences \cite{Wang12}.  We begin by showing how to perform universal, robust, single-qubit gates.  We then demonstrate how to use these single-qubit pulse sequences to generate a {\sc cnot} gate that possesses the same resilience against errors.  Finally, we show how to combine the prior two results to generate universal, multi-qubit, dynamically corrected operations on a large-scale quantum register.  Thus, our work forms a complete prescription for compensating low-frequency noise in ongoing singlet-triplet quantum computation experiments.

The model Hamiltonian within the logical subspace of a singlet-triplet qubit is written in terms of the Pauli operators $\mathbf{\sigma}$ as
\begin{equation}\label{eq:ham}
H(t)=\frac{h}{2}\sigma_x+\frac{J\left(\epsilon\left(t\right)\right)}{2}\sigma_z,
\end{equation}
with $h=g\mu_B\Delta B_z$ the energy associated with the average magnetic field gradient across the double-dot and $J$ the positive, bounded exchange coupling.  The exchange is a function of the energy difference between balanced and imbalanced singlet charge states, $\epsilon$, which can be controlled dynamically \cite{Petta05}.  Since both Overhauser and charge noise are typically several orders of magnitude slower than gate times, the resulting perturbations about $h$ and $\epsilon\left(t\right)$, $\delta h$ and $\delta \epsilon$, respectively, are treated as random constants.

A single-qubit rotation of angle $\phi$ about axis $h\hat{x}+J\hat{z}$, $R\left(h\hat{x}+J\hat{z},\phi\right)$, na\"{i}vely performed by holding the exchange constant over some time interval, results in errors $\Delta_i$ (see Supplement for the explicit formulas \cite{note}),
\begin{multline}\label{eq:UJphi}
U\left(J,\phi\right) \equiv \exp{\left[-i\left(\frac{h+\delta h}{2} \sigma_x + \frac{J+\delta J}{2} \sigma_z\right)\frac{\phi}{\sqrt{h^2+J^2}}\right]}
\\
= \exp{\left[-i\left(\frac{h}{2} \sigma_x + \frac{J}{2} \sigma_z\right)\frac{\phi}{\sqrt{h^2+J^2}}\right]}\left(I - i\sum_i\Delta_i \sigma_i \right).
\end{multline}
Here $h$ and $\delta h$ are assumed to be independent of $J$, and $\delta J = \delta\epsilon \frac{\partial J\left(\epsilon\right)}{\partial \epsilon}|_{J\left(\epsilon\right)=J}$ arises from fluctuations in the background impurity potential and hence in detuning, $\delta\epsilon$.

Our general strategy is to construct an identity operation such that the error in its implementation exactly cancels the leading order error in the original rotation.  For example, a $2\pi$ rotation interrupted by a $2\pi$ rotation about a different axis (the {\sc supcode} identity of Ref.~\cite{Wang12}) allows three degrees of freedom with which to tune the error: the two axes and the point of interruption.  Here we use a similar concept in a more general form to compensate for all error sources.  First, note that a $2m_n\pi$ rotation interrupted by a general zeroth order identity gives a new zeroth order identity with different first order errors,
\begin{multline}\label{eq:level1}
U\left(j_n,m_n\pi + \theta_n\right) \left(I - i \sum_i \delta^{\left(n-1\right)}_i \sigma_i \right) U\left(j_n,m_n\pi - \theta_n\right)
\\
= I - i \sum_i \delta^{\left(n\right)}_i \sigma_i
\end{multline}
This defines a recursion relation for the error of a ``level-$n$" parameterized identity,
\begin{multline}\label{eq:leveln}
U\left(j_n,m_n\pi + \theta_n\right)...U\left(j_1,m_1\pi + \theta_1\right) U\left(j_0,2m_0\pi\right)
\\
\times U\left(j_1,m_1\pi - \theta_1\right)...U\left(j_n,m_n\pi - \theta_n\right).
\end{multline}
The recursion equations are straightforward to generate \cite{note}, but their explicit form is lengthy and unnecessary for the discussion here.

We find that a level-5 identity contains enough flexibility to obtain errors which exactly cancel those of the na\"{i}ve rotation $U\left(J,\phi\right)$ to leading order in both $\delta h$ and $\delta\epsilon$.  This is the central new result of our work, and we choose a simple form for this compensating identity,
\begin{multline}\label{eq:correctedUJphi}
\!\!\!\!\!\!U\!\left(\!J,\pi+\frac{\phi}{2}\right)\! U\left(j_4,\pi\right) U\left(j_3,\pi\right) U\left(0,\pi\right) U\left(j_1,\pi\right) U\left(j_0,4\pi\right)
\\
\times U\left(j_1,\pi\right) U\left(0,\pi\right) U\left(j_3,\pi\right) U\left(j_4,\pi\right) U\left(J,\pi+\frac{\phi}{2}\right)
\\
= \exp{\left[-i\left(\frac{h}{2} \sigma_x + \frac{J}{2} \sigma_z\right)\frac{\phi}{\sqrt{h^2+J^2}}\right]} + \mathcal{O}\left[\left(\delta h+\delta\epsilon\right)^2\right].
\end{multline}
Here we have taken $j_5=J$, $\theta_{1,2,3,4}=0$, and $\theta_5=-\phi/2$ so that the overall sequence is symmetric and $\Delta_y+\delta^{\left(5\right)}_y=0$. We have arbitrarily set $j_2=0$ for simplicity.  The remaining parameters ${j_0,j_1,j_3,j_4}$ are determined by numerical solution of the four coupled, nonlinear equations that set the coefficients of $\delta h$ and $\delta \epsilon$ in both $\Delta_x+\delta^{\left(5\right)}_x$ and $\Delta_z+\delta^{\left(5\right)}_z$ to be zero \cite{note}.

\begin{figure}
  \includegraphics[width=0.9\columnwidth]{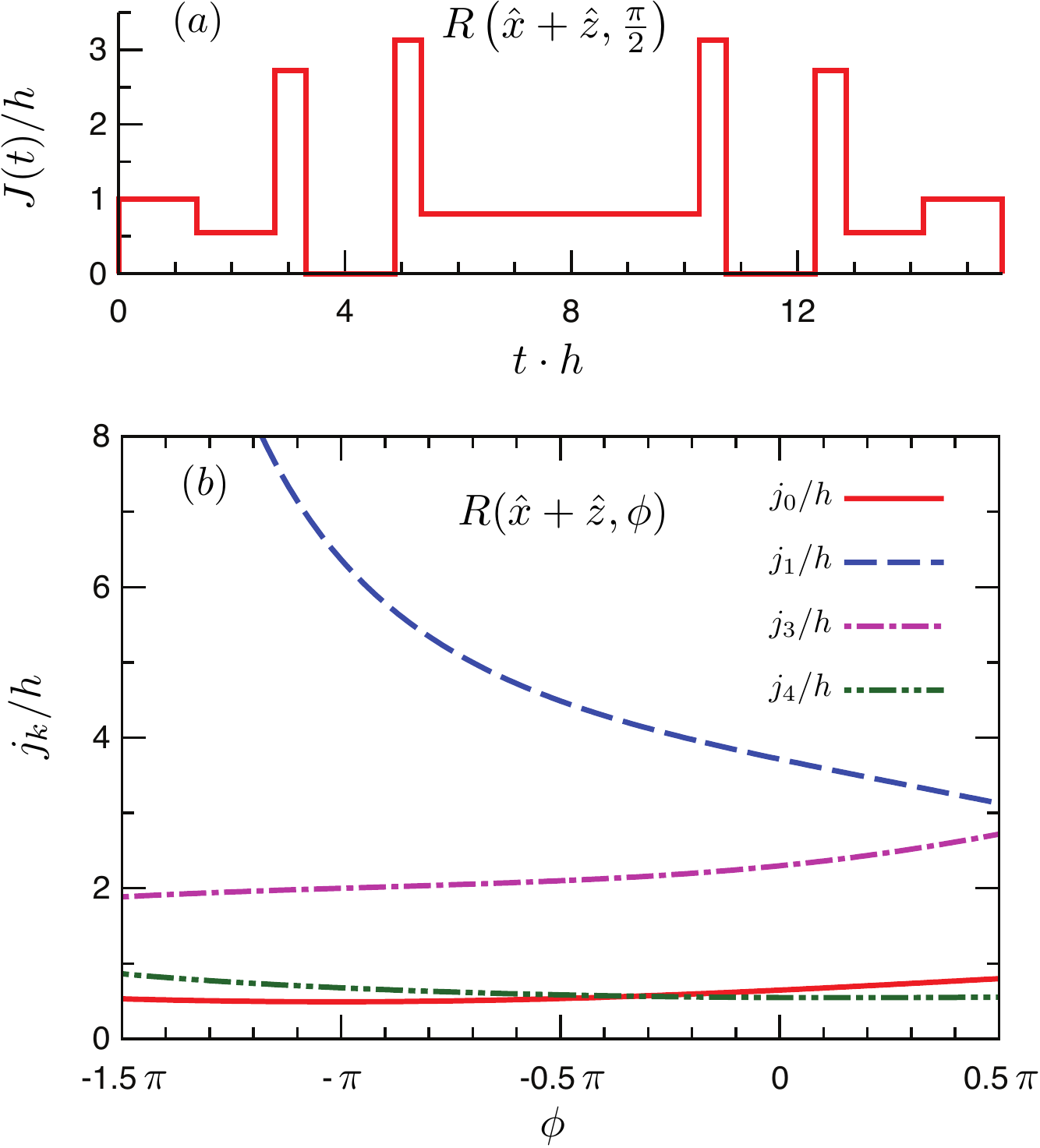}\\
  \caption{(Color online.) Rotations about $\hat{x}+\hat{z}$. (a) Example pulse sequence for $\pi/2$ rotation. (b) Parameters vs angle, $\phi$.}\label{fig:pulseshape}
\end{figure}
\begin{figure}
  \includegraphics[width=0.9\columnwidth]{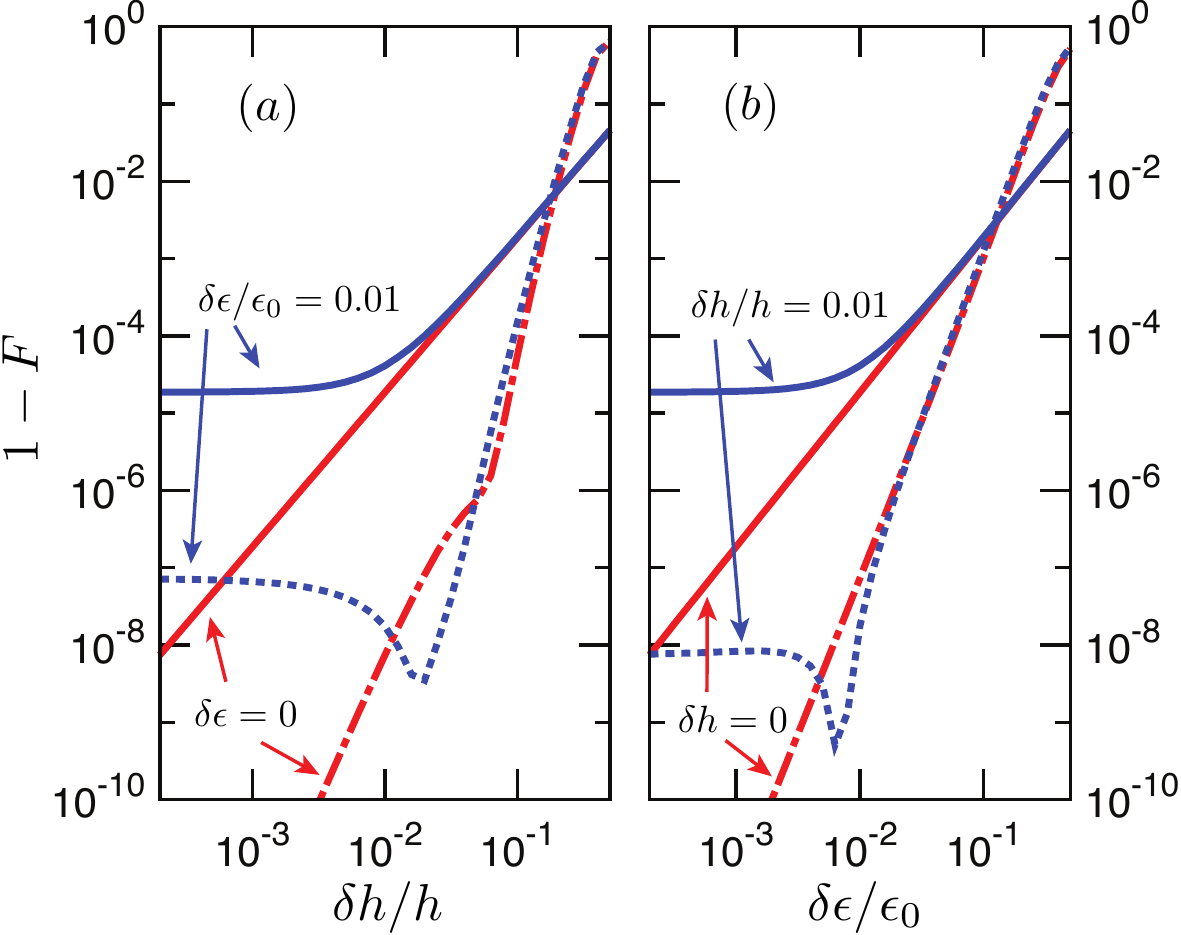}\\
  \caption{(Color online.) Gate infidelity of na\"{i}ve (solid) and corrected (broken) $\pi/2$ rotations about $\hat{x}+\hat{z}$ vs (a) Overhauser field gradient fluctuations, $\delta h/h$, and (b) detuning fluctuations, $\delta \epsilon/\epsilon_0$.}\label{fig:error}
\end{figure}

An example of this sequence, along with the dependence of the parameters on the rotation angle, is shown in Fig.~\ref{fig:pulseshape}.  While there are multiple possible solutions, we have constrained our numerical solution to give only positive values of the exchange coupling within a typically experimentally accessible region. In all cases, our restrictions to physically practical sequences still permit error compensation.
The suppression of error is evident in Fig.~\ref{fig:error}.  The corrected sequence infidelity scales as the fourth (instead of second) power of the fluctuations.  (Hence, if one reinterprets $\delta h$ and $\delta \epsilon$ as standard deviations of Gaussian distributions, the plotted corrected sequence infidelities should be multiplied by a factor of three.)  For fluctuations on the order of a few percent, our pulse sequence suppresses gate error by two orders of magnitude.

The results shown here are for the usual empirical model at negative $\epsilon$, $J\left(\epsilon\right) \propto J_0 + J_1 \exp\left(\epsilon/\epsilon_0\right)$ with $J_0 \sim 0$ so $\delta J/J = \delta\epsilon/\epsilon_0$ \cite{Shulman12,Dial12}.  However, our method does not require this assumption, and is equally valid for other models of $J\left(\epsilon\right)$, as we have also verified explicitly, including models where the exchange cannot be tuned all the way to zero.  The important requirement is simply that there \emph{is} a model so that the correlations between the $\delta J$s at different values of $J$ are known.  Also, although we have taken the pulse sequence to be piecewise constant, we have checked that including a finite rise time only introduces a perturbation about the parameter values plotted.  In an experimental context, then, an optimization of the actual parameters around the ideal pulse described here should quickly converge to the adjusted sequence for that particular real setup.
\begin{figure*}[t]
  \includegraphics[width=\linewidth]{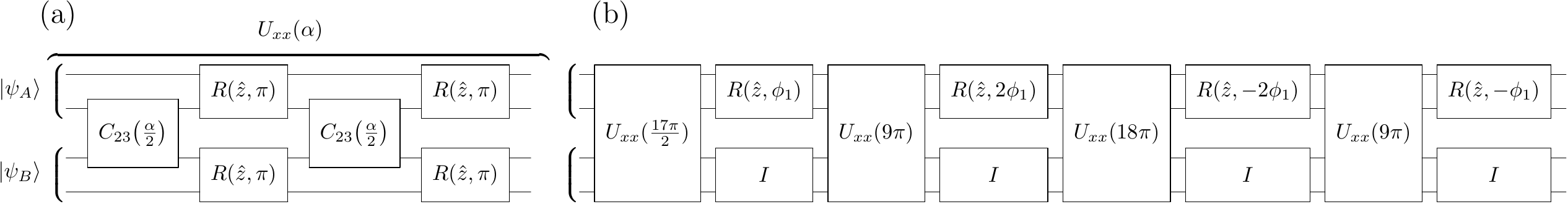}\\
  \caption{Quantum circuits for (a) Overhauser-corrected and (b) fully-corrected two-qubit Ising gates. Each block represents a rotation carried out by a composite pulse.}\label{fig:circuit}
\end{figure*}

The particular realization of the identity above is not a unique solution -- we had more parameters than equations and simply chose values for which solutions were easy to find.  Further optimization may yield an even shorter solution.  However, this identity uses only $14\pi$ of total rotation and is already quite efficient. (Note that an $\hat{x}+\hat{z}$ rotation with the original {\sc supcode} sequence \cite{Wang12}, only correcting $\delta h$, na\"{i}vely required $120\pi$ of total rotation in the identities, making the current protocol more than an order of magnitude more efficient than the original.)

So far, we have only discussed correcting rotations about axes of the form $h\hat{x}+J\hat{z}$.  Given the experimental constraints on $h$ and $J$, this only allows direct access to a segment of the positive $x$-$z$ quarter-plane. Clearly, one can build an arbitrary corrected single-qubit rotation from a string of the corrected rotations discussed above.  More efficiently, though, one could also build it from a string of uncorrected rotations about the accessible axes and then correct the total error with a single compensating identity rather than correcting each segment individually.  We have found that more general rotations can be corrected using a level-6 identity \cite{note}.

Now we turn to the construction of corrected two-qubit gates on a pair of singlet-triplet qubits.  Two-qubit gates have been demonstrated experimentally via capacitive coupling \cite{Shulman12}, and proposed theoretically via exchange coupling \cite{Li12,Klinovaja12}.  We will consider the latter case with neighboring qubits $A$, consisting of dots 1 and 2, and $B$, consisting of dots 3 and 4. The qubits interact via an exchange link between dots 2 and 3. In recent work, Ising gates are constructed in the absence of noise such that phases due to static local Overhauser fields cancel, and only a state-dependent phase from the exchange pulse survives \cite{Klinovaja12},
\begin{multline}\label{eq:Klinovaja}
U_{xx}\left(\alpha\right) \equiv R^{\left(A\right)}\left(\hat{z},\pi\right) R^{\left(B\right)}\left(\hat{z},\pi\right) C_{23}\left(\frac{\alpha}{2}\right)
\\
\times R^{\left(A\right)}\left(\hat{z},\pi\right) R^{\left(B\right)}\left(\hat{z},\pi\right) C_{23}\left(\frac{\alpha}{2}\right)
\\
= \exp\left(i\frac{\alpha}{2} \sigma_x^{\left(A\right)}\sigma_x^{\left(B\right)}\right) + \mathcal{O}\left(\delta h, \delta\epsilon\right),
\end{multline}
where $R^{\left(A/B\right)}\left(\hat{r},\theta\right)$ denotes a rotation of qubit $A/B$ by $\theta$ about $\hat{r}$ on the singlet-triplet Bloch sphere.  $C_{23}\left(\alpha/2\right)$ denotes the application of a pulse to the inter-qubit link such that in an $S_z=0$ subspace it would act as a $2\pi$ rotation about some axis.  This is required to avoid swapping anti-aligned spins on dots 2 and 3, which would cause leakage out of the logical qubit subspace.  The choice of axis is determined by fixing the total inter-qubit pulse area, $\int dt J_{23}\left(t\right) = \alpha/2$, which causes the desired relative phase to be acquired by two-qubit states with aligned electron spins on dots 2 and 3.

By using our compensated pulse sequences to perform the $R^{\left(A/B\right)}\left(\hat{z},\pi\right)$ operations -- and performing them on $A$ and $B$ with equal durations such that both qubits are idle for the same amount of time -- single-qubit errors on the rhs of Eq.~\eqref{eq:Klinovaja} are eliminated to first order.  For equal gradients on both qubits, the equal time condition is trivially satisfied.  More generally, the $\pi$ rotations on $A$ and $B$ would have different durations and one would pad them with corrected $2\pi$ rotations about different axes (i.e., having different durations) to compensate \cite{note}.

Also implementing the $C_{23}\left(\alpha/2\right)$ with our pulse sequences suppresses leakage error, but the angle $\alpha$ is still sensitive to charge noise induced exchange amplitude error.  This is because our sequences only correct rotations within the $S_z=0$ subspace of the $SU(2)^{\otimes 2}$ space on which $C_{23}$ acts.  The problem at this stage, though, is one familiar from NMR contexts -- pulse length error in Ising-coupled spins \cite{Jones03}. Defining a tilted version of the Ising gate,
\begin{equation}\label{eq:tiltedKlinovaja}
U_{xx}'\left(\alpha,\phi\right) \equiv  R^{\left(A\right)}\left(\hat{z},-\phi\right) U_{xx}\left(\alpha\right) R^{\left(A\right)}\left(\hat{z},\phi\right),
\end{equation}
the inter-qubit exchange amplitude error can be corrected using a BB1 sequence \cite{Wimperis94,Jones03}.  Typically we only report angles modulo $2\pi$, but because the error is proportional to $\alpha$ and using our composite pulses results in a large angle $\alpha = 8\pi + \text{mod}\left(\alpha,2\pi\right)$, we must use a slightly modified version of the BB1 sequence:
\begin{multline}\label{eq:BB1}
U_{xx}'\left(\theta_1,\phi_1\right) U_{xx}'\left(2\theta_1,3\phi_1\right) U_{xx}'\left(\theta_1,\phi_1\right) U_{xx}\left(\alpha\right)
\\
= \exp\left(i\frac{\alpha}{2} \sigma_x^{\left(A\right)}\sigma_x^{\left(B\right)}\right) + \mathcal{O}\left[\left(\delta h + \delta\epsilon \right)^2\right],
\end{multline}
where $\theta_1 = 8\pi+\pi$, $\phi_1 = \arccos{\left(-\alpha/4\left(8\pi+\pi\right)\right)}$ \cite{bb1note}, and all rotations are corrected composites.  For $\alpha = 8\pi+\pi/2$, this gate is equivalent to a {\sc cnot} gate up to single-qubit operations \cite{Li12,Klinovaja12}.  In practice, we also perform corrected identities on qubit $B$ during the rotations on $A$ in Eq.~\eqref{eq:tiltedKlinovaja} to avoid reintroducing single-qubit errors.

\begin{figure}
  \includegraphics[width=0.9\columnwidth]{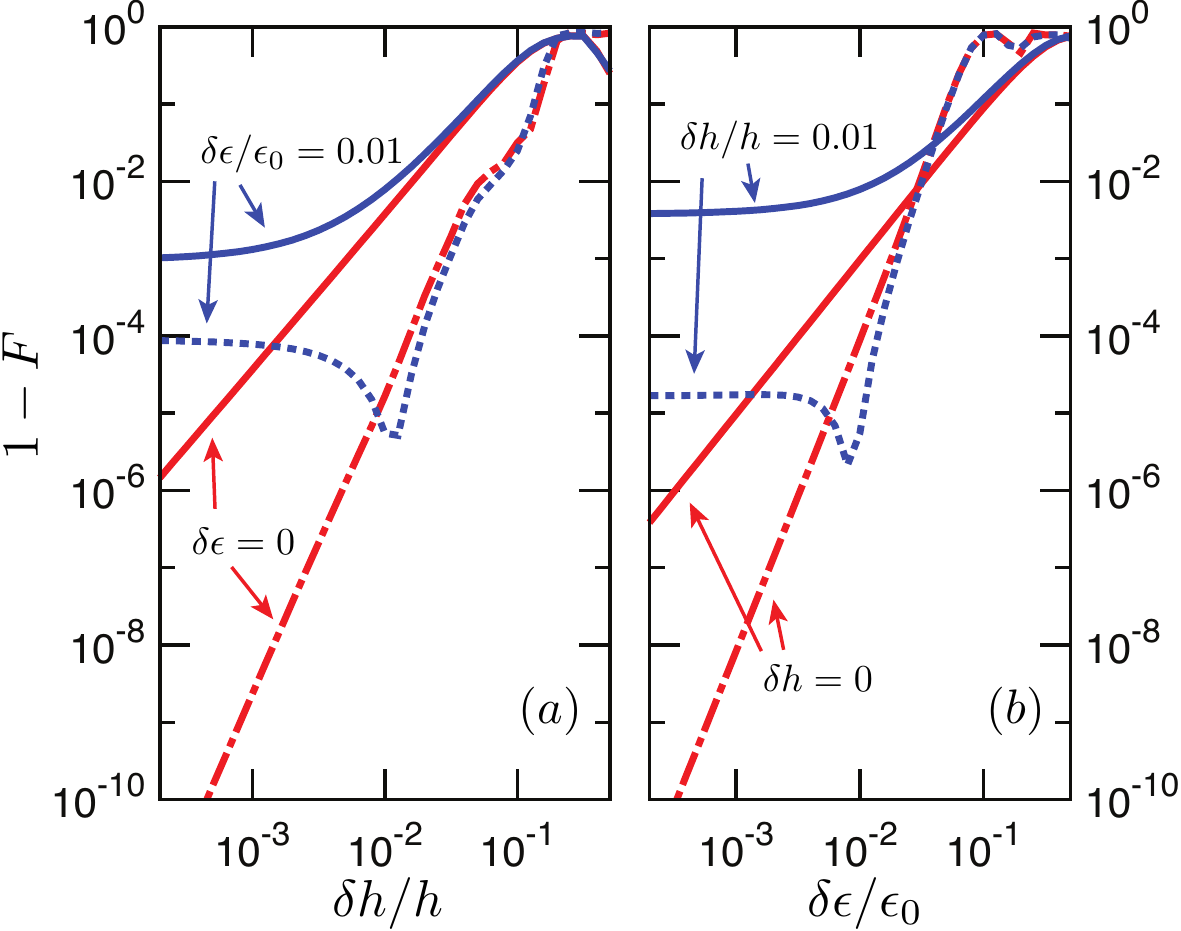}\\
  \caption{(Color online.) Gate infidelity \cite{note} of uncorrected (solid) and corrected (dashed) {\sc cnot} operations vs (a) Overhauser field gradient fluctuations, $\delta h/h$, and (b) detuning fluctuations, $\delta \epsilon/\epsilon_0$.}\label{fig:CNOTerror}
\end{figure}

We sketch the implementation of the Ising $\pi/2$ gate in Fig.~\ref{fig:circuit}.  We have displayed the case of a linear gradient, after trivial contractions of sequential single-qubit operations in Eq.~\eqref{eq:BB1}. Each line represents a quantum dot, which may be linked to its neighbor by an exchange pulse.  A singlet-triplet qubit then is denoted by a pair of lines which are understood to have total spin projection zero.
The error suppression for a {\sc cnot} gate is shown in Fig.~\ref{fig:CNOTerror}.  For the sake of display, we have reduced the number of parameters by assuming a uniformly fluctuating linear field gradient across the four dots, as well as identical detuning fluctuations on each link.

As is always the case with composite pulse schemes, the cost of the correction is longer gate time.  Currently, our shortest implementation of the corrected {\sc cnot} gate requires 20 composite pulses, corresponding to $\sim 300\pi$ of total rotation.  While certainly challenging, this is well within the realm of possibility given that our sequence suppresses decoherence during its operation.  Recall that statistical fluctuations of the control Hamiltonian lead to rapid decoherence on the free induction time scale, $T_2^{\ast}$.  Dynamical decoupling can extend coherence to a much longer time $T_2$, though, where dynamical fluctuations become important.  Roughly speaking, gate error scales like the ratio of gate time to coherence time.  Since typically $T_2 \geq 10^4 T_2^{\ast}$ in semiconductor spin systems near the charge degeneracy point \cite{Petta05,Bluhm11,Maune12,Witzel10}, if one works in that regime it is well worth using longer gate implementations that simultaneously perform dynamical decoupling such that the relevant coherence time is close to $T_2$ rather than $T_2^{\ast}$.  (Although note that if the qubit is operated in a region where $J$ is more sensitive to $\epsilon$, as done intentionally in Ref.~\cite{Shulman12}, $T_2$ will decrease in the absence of charge noise correction.)  Stated differently, the ratio of the timescale on which dynamical fluctuations can be expected to the time duration of a simple $\pi$ rotation can be roughly estimated as $\sim 10^6$, so it is logical to sacrifice some rotation time to gain additional precision.

Since each experiment will have its own preferred region of control space, further optimization would be premature, but future work will almost certainly reduce the length in any case.  Also, it may well be possible to construct a ``one-shot" correction for the {\sc cnot} gate, as we have done for the composite single-qubit rotations, rather than correcting at each intermediate stage.  These extensions, however, are beyond the scope of the present work.

Finally, we discuss application to an array of exchange-coupled qubits.  Our method facilitates operations below the quantum error correction threshold, a vital requirement for scaling up.  Arbitrary corrected multi-qubit circuits can be implemented similarly to Fig.~\ref{fig:circuit} as long as there exist single-qubit corrected identities of sufficiently variable duration to protect the many idle qubits during the application of nontrivial gates to other qubits.  We find that level-6 identities can be used to cover the entire range of relevant idling times \cite{note}.  An additional benefit of using our sequences in large systems is that errors due to small spatial inhomogeneities in the control Hamiltonian \eqref{eq:ham} are also suppressed.

In summary, we have shown above that the relevant types of error in experimental singlet-triplet spin qubit manipulation can be eliminated to leading order by a new composite pulse sequence.  We show how to apply this sequence for single-qubit, two-qubit, and multi-qubit systems, opening a path towards scalable, universal quantum computation in a noisy solid-state environment.

Research was supported by LPS and by the Office of the Director of National Intelligence, Intelligence Advanced Research Projects Activity (IARPA), through the Army Research Office grant W911NF-12-1-0354.


\onecolumngrid

\vspace{1cm}
\begin{center}
{\bf\large Supplementary material}
\end{center}
\vspace{0.5cm}

\setcounter{secnumdepth}{3}  
\setcounter{equation}{0}
\setcounter{figure}{0}
\renewcommand{\theequation}{S-\arabic{equation}}
\renewcommand{\thefigure}{S\arabic{figure}}
\renewcommand\figurename{Supplementary Figure}
\newcommand\Scite[1]{[S\citealp{#1}]}

\makeatletter \renewcommand\@biblabel[1]{[S#1]} \makeatother


\section{\texorpdfstring{Error terms in $\boldsymbol{U(J,\phi)}$ and uncorrected identity}{Error terms in U(J,phi) and uncorrected identity}}

In this section we present the explicit form of the error terms in $U(J,\phi)$, which is defined in Eq.~\eqref{eq:UJphi}.

We start with the error terms in Eq.~\eqref{eq:UJphi}, which, to the first order of $\delta h$ and $\delta J$, are
\begin{subequations}
\begin{align}
  \Delta_x &= \delta h \frac{h^2\phi + J^2\sin{\phi}}{2\left(h^2+J^2\right)^{3/2}} + \delta J \frac{h J\left(\phi-\sin{\phi}\right)}{2\left(h^2+J^2\right)^{3/2}}, \\
  \Delta_y &= \delta h \frac{J\left(\cos{\phi}-1\right)}{2\left(h^2+J^2\right)} + \delta J \frac{h\left(1-\cos{\phi}\right)}{2\left(h^2+J^2\right)}, \\
  \Delta_z &= \delta h \frac{h J\left(\phi-\sin{\phi}\right)}{2\left(h^2+J^2\right)^{3/2}} + \delta J \frac{\left(J^2\phi + h^2\sin{\phi}\right)}{2\left(h^2+J^2\right)^{3/2}}.
\end{align}
\end{subequations}

We aim to cancel these first-order error by supplementing the rotation with a level-$n$ uncorrected identity $\widetilde{I}$,
\begin{align}
\widetilde{I}^{(n)}=I - i\sum_{i}\delta_i^{(n)} \sigma_i,\label{suppleq:recursident}
\end{align}
such that $\Delta_i+\delta_i^{(n)}=0$ for $i=x,y,z$.

The uncorrected identity we use in our work is defined recursively in Eqs.~\eqref{eq:level1} and \eqref{eq:leveln}. It is therefore useful to establish the recursion relation for the coefficients. At zeroth order, $\widetilde{I}^{(0)}$ is simply $U\left(j_0,2m_0\pi\right)$, therefore
\begin{subequations}
\begin{align}
\delta^{\left(0\right)}_x &= \frac{m_0\pi h\left(h\delta h + j_0\delta j_0\right)}{\left(h^2+j_0^2\right)^{3/2}}, \label{suppleq:bc1}
\\
\delta^{\left(0\right)}_y &= 0,\label{suppleq:bc2}
\\
\delta^{\left(0\right)}_z &= \frac{m_0\pi j_0\left(h\delta h + j_0\delta j_0\right)}{\left(h^2+j_0^2\right)^{3/2}}. \label{suppleq:bc3}
\end{align}
\end{subequations}

According to Eqs.~\eqref{eq:level1} and \eqref{eq:leveln}, given $j_n$, $m_n$ and $\theta_n$, $\delta_i^{(n)}$ can be generated recursively as
\begin{subequations}
\begin{align}
\delta^{\left(n\right)}_x &= \frac{m_n\pi h\left(h\delta h + j_n\delta j_n\right)}{\left(h^2+j_n^2\right)^{3/2}} + \delta^{\left(n-1\right)}_x \frac{h^2 + \left(-1\right)^{m_n} j_n^2\cos{\theta_n}}{h^2+j_n^2} + \delta^{\left(n-1\right)}_y \left(-1\right)^{m_n} \frac{j_n\sin{\theta_n}}{\sqrt{h^2+j_n^2}}\notag\\
 &\quad+ \delta^{\left(n-1\right)}_z \frac{hj_n\left(1-\left(-1\right)^{m_n} \cos{\theta_n}\right)}{h^2+j_n^2}, \label{suppleq:recursion1}
\\
\delta^{\left(n\right)}_y &= \left(-1\right)^{m_n} \left(-\delta^{\left(n-1\right)}_x \frac{j_n\sin{\theta_n}}{\sqrt{h^2+j_n^2}} + \delta^{\left(n-1\right)}_y \cos{\theta_n} + \delta^{\left(n-1\right)}_z \frac{h\sin{\theta_n}}{\sqrt{h^2+j_n^2}} \right), \label{suppleq:recursion2}
\\
\delta^{\left(n\right)}_z &= \frac{m_n\pi j_n\left(h\delta h + j_n\delta j_n\right)}{\left(h^2+j_n^2\right)^{3/2}} + \delta^{\left(n-1\right)}_x \frac{hj_n\left(1-\left(-1\right)^{m_n} \cos{\theta_n}\right)}{h^2+j_n^2} - \delta^{\left(n-1\right)}_y \left(-1\right)^{m_n} \frac{h\sin{\theta_n}}{\sqrt{h^2+j_n^2}}\notag\\
&\quad+ \delta^{\left(n-1\right)}_z \frac{j_n^2 +\left(-1\right)^{m_n} h^2\cos{\theta_n}}{h^2+j_n^2}. \label{suppleq:recursion3}
\end{align}
\end{subequations}

For convenience, from here onward we take $h=1$ as our energy unit. We also denote $g(J)=\frac{\partial J\left(\epsilon\right)}{\partial \epsilon}|_{J\left(\epsilon\right)=J}$ so that $\delta J = g(J)\delta\epsilon$.

\section{\texorpdfstring{Rotation around axis $\boldsymbol{\hat{x}+J\hat{z}}$}{Rotation around axis x+J z}}\label{supplsec:onepiece}

This section discusses how one can correct a rotation of angle $\phi$ around an axis determined by $\hat{x}+J\hat{z}$, $U(J,\phi)$.

According to the main text [see Eq.~\eqref{eq:correctedUJphi}], our target is to design an uncorrected identity $\widetilde{I}$, satisfying
\begin{align}
U\left(J,\phi\right)\widetilde{I}=e^{i\chi}R(\hat{x}+J\hat{z},\phi)
\left\{I+{\cal O}\left[\left(\delta h+\delta\epsilon\right)^2\right]\right\}
\label{suppleq:one-piece}
\end{align}
where $\chi$ is an unimportant phase factor.

We correct this rotation using a level-5 identity. We have also chosen the pulse to be symmetric, namely $\theta_5=-\phi/2$ and $\theta_{1,2,3,4}=0$. In this case, the $\sigma_y$ term would automatically vanish in the final rotation (see the ``Methods'' section of Ref.~\cite{Wang12}), and we only need to solve four coupled nonlinear equations.

We can absorb the outmost level into $U\left(J,\phi\right)$ and are left with a level-4 identity whose parameters are fixed by the conditions
\begin{subequations}
\begin{align}
\delta_x^{(4)}&=-\frac{2 \pi+\phi-2 J^2 \sin\frac{\phi}{2}}{2 \left(1+J^2\right)^{3/2}}\delta h-\frac{J g(J) \left(2 \pi +\phi +2 \sin\frac{\phi}{2}\right)}{2 \left(1+J^2\right)^{3/2}}\delta J,\\
\delta_z^{(4)}&=-\frac{J \left(2 \pi +\phi +2 \sin\frac{\phi}{2}\right)}{2 \left(1+J^2\right)^{3/2}}\delta h-\frac{g(J) \left[J^2 (2 \pi +\phi )-2 \sin\frac{\phi}{2}\right]}{2 \left(1+J^2\right)^{3/2}}\delta J.\label{suppleq:onepieceeq4}
\end{align}
\end{subequations}

As explained in the main text, we can construct the identity as
\begin{align}
\widetilde{I}^{(4)}=U\left(j_4,\pi\right) U\left(j_3,\pi\right) U\left(j_2,\pi\right) U\left(j_1,\pi\right) U\left(j_0,4\pi\right)U\left(j_1,\pi\right) U\left(j_2,\pi\right) U\left(j_3,\pi\right) U\left(j_4,\pi\right).
\end{align}
Since we only have four equations, we only need four parameters to determine. Here we choose $j_2=0$. (We have also verified that our method does not require this assumption to work, namely one can choose a different value of $j_2$, or one can instead fix another variable,  e.g. $j_1$.)

\section{\texorpdfstring{Rotation around $\boldsymbol{\hat{z}}$-axis}{Rotation around z-axis}}\label{supplsec:zrot}

The zeroth order $\hat{z}$-axis rotation, $R(\hat{z},\phi)$, is achieved by \cite{Wang12}:
\begin{align}
R(\hat{x}+\hat{z},\pi)R(\hat{x},\phi)R(\hat{x}+\hat{z},\pi).\label{suppleq:zrotzerothorder}
\end{align}

At first glance, since each of the three terms in Eq.~\eqref{suppleq:zrotzerothorder} can be implemented using the pulse of Sec.~\ref{supplsec:onepiece}, it is straightforward to implement a dynamically corrected $z$-rotation. However, this would require $40\pi\sim50\pi$ of sweeps around the Bloch sphere, we are interested in further optimizing it. We shall try to do a ``one-shot'' correction, namely correcting the pulses of Eq.~\eqref{suppleq:zrotzerothorder} with only one identity.

One trick to be employed here is that we shall insert the identity between $R(\hat{x},\phi)$ and $R(\hat{x}+\hat{z},\pi)$ but not at the right end of Eq.~\eqref{suppleq:zrotzerothorder}, so the corrected pulse looks like
\begin{align}
U(J=1,\pi)U(J=0,\phi)\widetilde{I}^{(6)}U(J=1,\pi).
\end{align}
As in Sec.~\ref{supplsec:onepiece}, we absorb the outmost level of that identity into $U(J=0,\phi)$ so that the corrected pulse is
\begin{align}
U(J=1,\pi)U\left(J=0,\pi+\frac{\phi}{2}\right)\widetilde{I}^{(5)}U\left(J=0,\pi+\frac{\phi}{2}\right)U(J=1,\pi).\label{suppleq:zrotexpand}
\end{align}
Therefore if we implement $\widetilde{I}^{(5)}$ in a symmetric way then the entire sequence is also symmetric, assuring us that the $\sigma_y$ component will automatically vanish and we are only left with four rather than six equations to solve. This is why we place the uncorrected identity in the middle of the sequence. In fact, an uncorrected identity operation can be placed anywhere and one may simply choose a location which is most convenient.

We will not explicitly carry out the algebra here, but we find the following sequence already accomplishes our purpose:
\begin{align}
\begin{split}
&U(J=1,\pi)
U(j_5=0,2\pi+\frac{\phi}{2})
U(j_4,\pi)
U(j_3,\pi)
U(j_2,\pi)
U(j_1=0,\pi)
U(j_0,4\pi)\\
&\times U(j_1=0,\pi)
U(j_2,\pi)
U(j_3,\pi)
U(j_4,\pi)
U(j_5=0,2\pi+\frac{\phi}{2})
U(J=1,\pi)
\end{split}\label{suppleq:zpulsecorr}
\end{align}
The parameters $j_{0,2,3,4}$ are given in Supplementary Fig.~\ref{supplfig:zpulse}. Note that for a small range of $\phi$ ($0.6\pi\sim0.9\pi$), $j_3$ becomes negative, making the solution unphysical. Although this problem can be easily overcome by duplicating pulses that we can physically do, it is more natural to stick to a ``one-shot'' correction to keep it optimal. In fact, we have explicitly verified that, setting $j_2=j_4=0$ (instead of $j_1=j_5=0$ here) while solving for $j_0$, $j_1$, $j_3$, $j_5$ will generate a physical solution to the pulse sequence covering this $\phi$ range.

We now consider the total rotation angle of the pulse sequence Eq.~\eqref{suppleq:zpulsecorr}. We see that it requires around $18\pi\sim20\pi$ of rotation on the Bloch sphere, and is more than a factor of two shorter than the na\"ive speculation at the beginning of this section.

\begin{figure}[]
    \centering
    \includegraphics[width=8cm, angle=0]{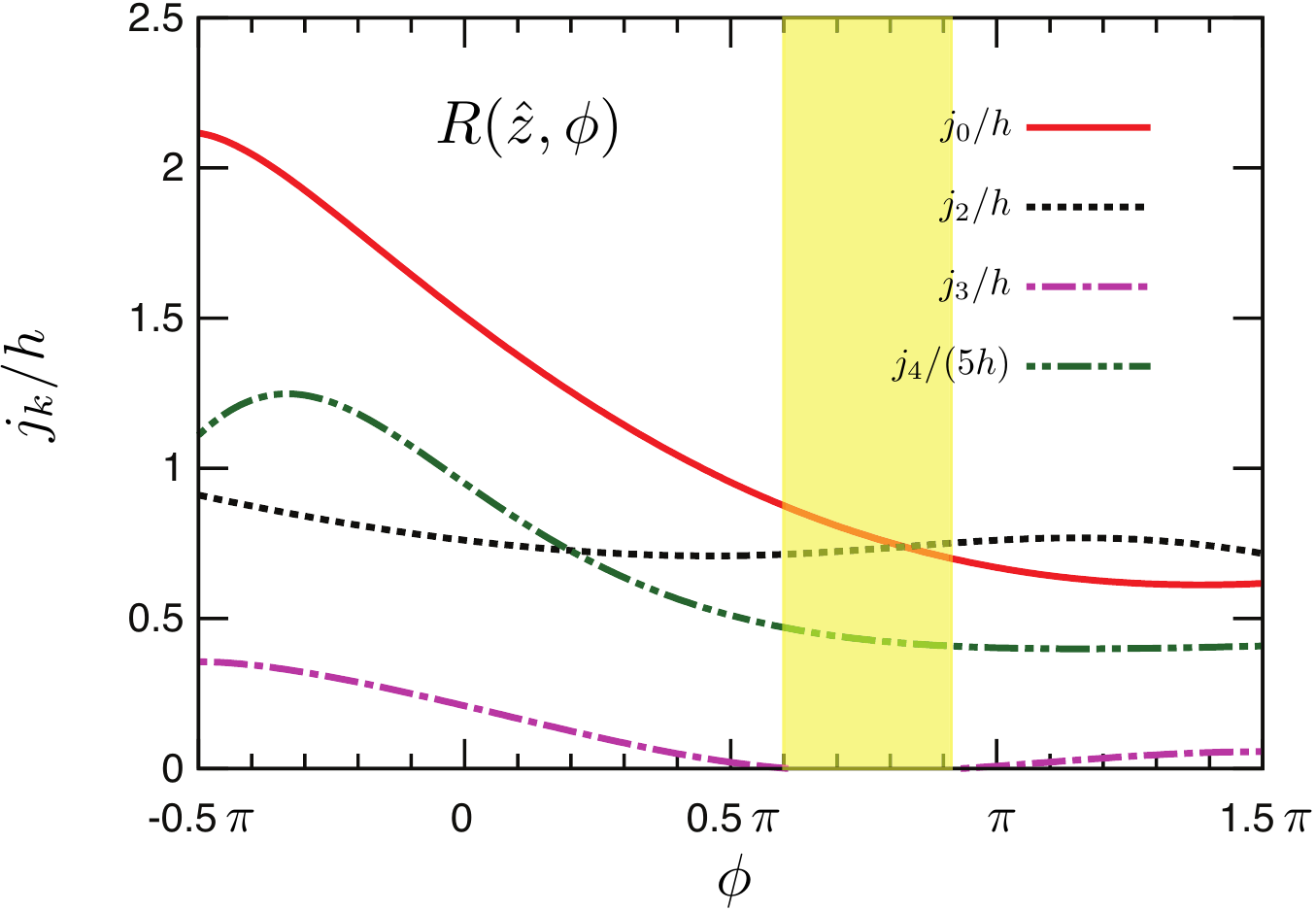}
    \caption{Parameters v.s. angle for rotations about $\hat{z}$. $g(J)=J/\epsilon_0$. The shaded area around $0.6\pi<\phi<0.9\pi$ indicates the range where $j_3$ becomes negative, rendering the solution unphysical.}
    \label{supplfig:zpulse}
\end{figure}

\section{Arbitrary rotation}

It is well-known that an arbitrary rotation can be decomposed as
\begin{align}
R(\hat{x},\phi_a)R(\hat{z},\phi_b)R(\hat{x},\phi_c).
\end{align}
As we already know from Sec.~\ref{supplsec:zrot}, this decomposition can be written as
\begin{align}
R(\hat{x},\phi_a)R(\hat{x}+\hat{z},\pi)R(\hat{x},\phi_b)R(\hat{x}+\hat{z},\pi)R(\hat{x},\phi_c)\label{suppleq:arbseqzeroth}
\end{align}
up to a phase factor. Again, although we have the ability to correct each term in Eq.~\eqref{suppleq:arbseqzeroth}, we would prefer a ``one-shot'' correction at the cost of introducing a slightly higher level of uncorrected identity operation.

We note that for an arbitrary rotation, $\sigma_y$ terms are in general present, so typically we have to solve all six equations.
The identity is inserted to the sequence as
\begin{align}
U(J=0,\phi_a)U(J=1,\pi)\widetilde{I}U(J=0,\phi_b)U(J=1,\pi)U(J=0,\phi_c).\label{suppleq:arbcorrschematic}
\end{align}
We still insert $\widetilde{I}$ in the middle of the sequence with the hope that when the rotation axis is on the $x$-$z$ plane but is not implementable using methods described in Secs.~\ref{supplsec:onepiece} and \ref{supplsec:zrot} (for example in the second quadrant of the plane, or in the first quadrant but very close to the $z$-axis), this entire pulse sequence can still be made symmetric so one is to solve four instead of six equations. In the work presented here we use the sequence in Eq.~\eqref{suppleq:arbcorrschematic} since we already find physical solutions for a vast range of rotations. However one should note that this is not necessary, and one can seat the uncorrected identity (or identities) at any position in the sequence as long as it provides sufficient degrees of freedom to find physical solution to the array of six coupled nonlinear equations.

We illustrate our sequence with a $R(\hat{y},\phi)$ rotation, which is an important operation used, for example, in converting the two-qubit Ising gate to a {\sc cnot} gate \cite{Klinovaja12,Li12}. Obviously this would have $\sigma_y$ terms in the final sequence and one need to solve six equations.  We therefore consider a level-6 identity with $\theta_6$ and $j_0$ through $j_6$ to be determined. (We take $j_2=0$ to keep the number of unknown variables six.) The pulse sequence for a corrected $y$-rotation is
\begin{align}
\begin{split}
&U(J=0,\phi_a=\frac{3\pi}{2})
U(J=1,\pi)
U(j_6,\pi-\theta_6)
U(j_5,\pi)
U(j_4,\pi)
U(j_3,\pi)
U(j_2=0,\pi)
U(j_1,\pi)
U(j_0,4\pi)\\
&\times U(j_1,\pi)
U(j_2=0,\pi)
U(j_3,\pi)
U(j_4,\pi)
U(j_5,\pi)
U(j_6,\pi+\theta_6)
U(J=0,\phi_b=\phi)
U(J=1,\pi)
U(J=0,\phi_c=\frac{\pi}{2}),
\end{split}\label{suppleq:zpulsecorrb}
\end{align}
and numerical results of $j_k$ and $\theta_6$ for a range of rotation angles are shown in Supplementary Fig.~\ref{supplfig:ypulsepara}. We remark that the pulse sequence Eq.~\eqref{suppleq:zpulsecorrb} sweeps a total angle of $20\pi\sim22\pi$ around the Bloch sphere. Comparing to a na\"ive correction of an $x$-$z$-$x$ sequence which would cost $40\pi\sim50\pi$, this is again a factor of two improvement.

\begin{figure}[]
    \centering
    \includegraphics[width=8cm, angle=0]{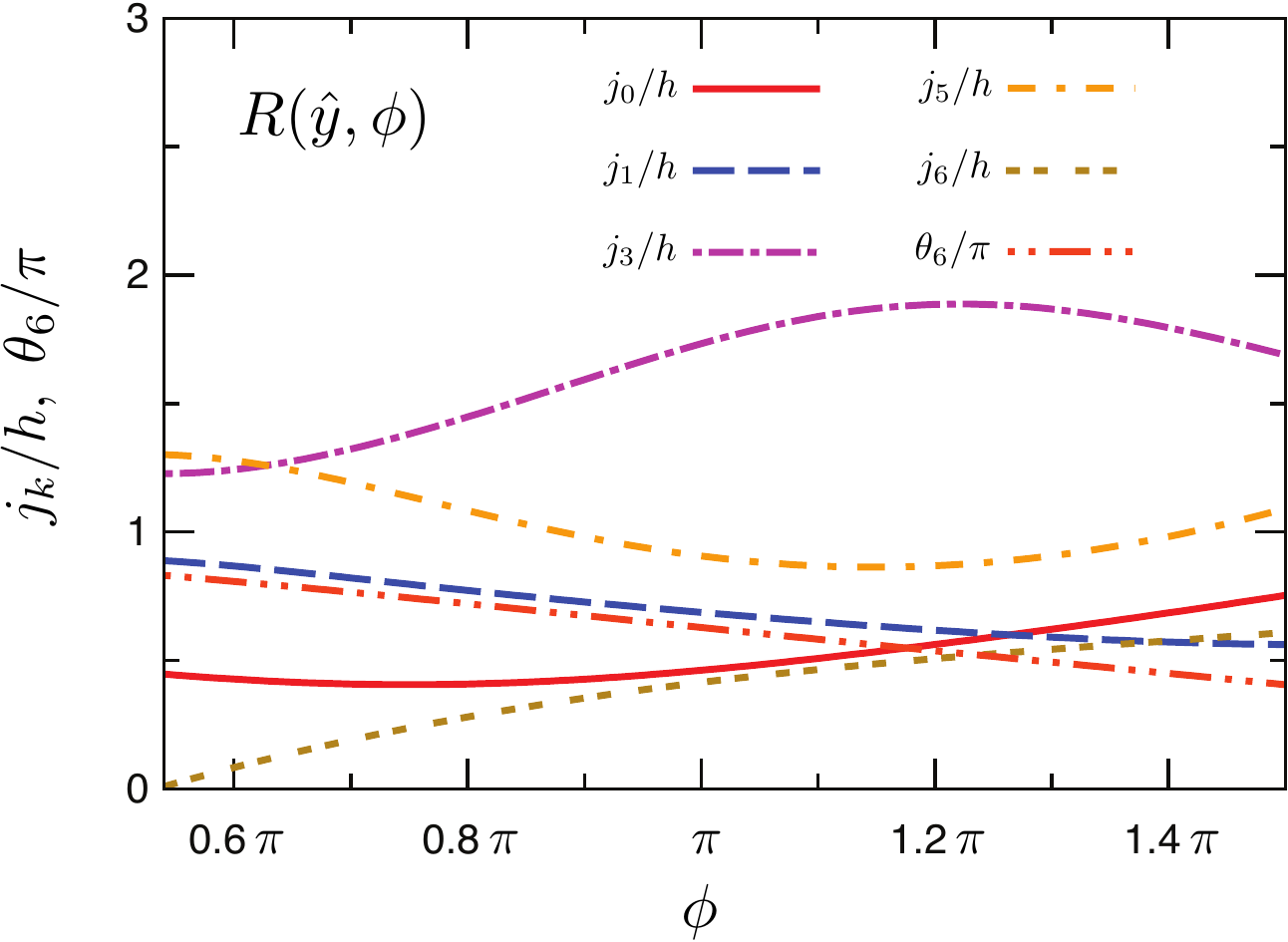}
    \caption{Parameters v.s. angle for rotations about $\hat{y}$. $g(J)=J/\epsilon_0$.}
    \label{supplfig:ypulsepara}
\end{figure}

\section{Identity operation}

The identity operation $\widetilde{I}$, sometimes called the {\sc noop}  operation, also plays a role in applying our method to two-qubit and multi-qubit system. First, at the two-qubit level it provides a fundamental ingredient of the entangled gate, defined as $C_{23}\left(\frac{\alpha}{2}\right)$ [cf. Eq.~\eqref{eq:Klinovaja} and Fig.~\ref{fig:circuit}(a)]. Here $\alpha$ is related to $\int dt J_{23}\left(t\right)$, which, in turn, determines the argument of the Ising gate $U_{xx}(\alpha)$.
Second, since the Overhauser field gradient is always present during the entire duration of performing a quantum algorithm, if only a subset of the multi-qubit system is being gated, the remaining qubits must carry out corrected identity operations in order to reduce vulnerability to noise. In this case, the timing is crucial. For example if one of the qubits is being rotated with time duration $T$, then the rest of the multi-qubit system must perform corrected identity operation with the same time duration $T$ [cf. Fig.~\ref{fig:circuit}(b)]. On the other hand, if one qubit is being gated with time $t_1$ while another $t_2$, then they must be supplemented by identities with time duration $T-t_1$ and $T-t_2$ respectively while others perform identity with time $T$.

These considerations mean that we must find identity operations which can generate a range of time duration as well as $\int dt J\left(t\right)$. We managed to do this with the following pulse sequence:
\begin{align}
\begin{split}
&U(J,2\pi)
U(j_5,\pi)
U(j_4=0,\pi)
U(j_3,\pi)
U(j_2=0,\pi)
U(j_1,\pi)
U(j_0,4\pi)\\
&\times U(j_1,\pi)
U(j_2=0,\pi)
U(j_3,\pi)
U(j_4=0,\pi)
U(j_5,\pi)
U(J,2\pi)
\end{split}\label{suppleq:identity}
\end{align}

The pulse sequence of Eq.~\eqref{suppleq:identity} generates identity operations with $T$ between $11/h$  and $24/h$ by varying $J$, from which identities with any desired time $T>11/h$ can be derived. We have used these identity operations in constructing the two-qubit gates schematically shown in Fig.~\ref{fig:circuit}(b). This pulse sequence at the same time generates the integrated value of $J(t)$, $\int dt J\left(t\right)$, between $10$ and $20$ (note that this corresponds to $20<\alpha<40$). Therefore in Fig.~\ref{fig:circuit}(b), $U_{xx}(17\pi/2)$ and $U_{xx}(9\pi)$ can be generated using this sequence. $U_{xx}(18\pi)$ can either be generated by duplicating $U_{xx}(9\pi)$, or, more optimally, be generated by a sequence with one higher level of identity operation, for which we shall not present details here.

\section{Definition of fidelity}

Figs.~\ref{fig:error} and \ref{fig:CNOTerror} show the performance of our pulse sequences in terms of the infidelity (one minus the fidelity, $F$). For a single qubit the fidelity is straightforwardly defined \Scite{Bowdrey02}. To quantify the fidelity of the two-qubit gate, we adopt the result of Ref.~\Scite{Cabrera07}. However, note that due to presence of the leakage subspaces in our system, our evolution operator is not trace-preserving. Therefore Eq.~(32) of Ref. \Scite{Cabrera07} must be extended as
\begin{equation}
 F= \frac{1}{16} \Biggl[ \frac{4}{5} \mathrm{Tr}\bigl[\sigma_0\otimes\sigma_0 U_f \sigma_0\otimes\sigma_0 U_f^\dag\bigr]+
    \frac{1}{5}\sum_{\substack{0\le\mu\le3\\0\le\nu\le3}}\mathrm{Tr}\bigl[V \sigma_\mu\otimes\sigma_\nu V^\dag U_f \sigma_\mu\otimes\sigma_\nu U_f^\dag\bigr]\Biggr] ,
\end{equation}
where $\sigma_0$ is the $2\times2$ identity matrix, and $X\otimes Y$ denotes an operator entirely within the computational subspace that acts as $X$ on the first qubit and as $Y$ on the second qubit (for example $\sigma_0\otimes\sigma_0$ is the projector into the computational subspace), $V$ is the desired operation with identity in the leakage subspace, and $U_f$ is the actual time evolution at the conclusion of the composite pulse sequence.




\begin{thebibliography}{99}
\bibitem{bb1note} These angles are for $J\left(\epsilon\right) \propto \exp\left(\epsilon/\epsilon_0\right)$, but more general models are easily accommodated.
\end{thebibliography}
\end{document}